\documentclass[pra,twocolumn,graphicx]{revtex4}
\usepackage{amssymb}
\usepackage{epsfig,amsmath}

\begin{document}

\title{Bound-state corrections in laser-induced nonsequential double
ionization}

\author{C. Figueira de Morisson Faria$^{1,2}$
 and M. Lewenstein$^{1,3}$\footnote{Also at: Insituci\'{o} Catalana di Recerca i Estudis
Avan\c{c}ats}} \affiliation{$^1$Institut f\"ur theoretische Physik,
Universit\"at Hannover, Appelstr. 2,
D-30167 Hannover, Germany\\
$^2$School of Engineering and Mathematical Sciences, City
University, Northampton Square, London EC1V OHB, United Kingdom\\
$^3$ ICFO - Institut de Ci\`{e}ncies Fot\`{o}niques, Jordi Girona
29, E-08034 Barcelona, Spain}
\date{\today}

\begin{abstract}
We perform a systematic analysis of how nonsequential double
ionization in intense, linearly polarized laser fields is influenced
by the initial states in which both electrons are bound, and by the
residual ionic potential. We assume that the second electron is
released by electron-impact ionization of the first electron with
its parent ion, using an S-Matrix approach. We work within the
Strong-Field Approximation, and compute differential electron
momentum distributions using saddle-point methods. Specifically, we
consider electrons in $1s$, $2p$, $3p$ and localized states, which
are released by either a contact or a Coulomb-type interaction. We
also perform an adequate treatment of the bound-state singularity
which is present in this framework. We show that the momentum
distributions are very sensitive with respect to spatially extended
or localized wave functions, but are not considerably influenced by
their shape. Furthermore, the modifications performed in order to
overcome the bound-state singularity do not significantly alter the
momentum distributions, apart from a minor suppression in the region
of small momenta. The only radical changes occur if one employs
effective form factors, which take into account the presence of the
residual ion upon rescattering. If the ionic potential is of contact
type, it outweighs the spreading caused by a long-range
electron-electron interaction, or by spatially extended bound
states. This leads to momentum distributions which exhibit a very
good agreement with the existing experiments.
\end{abstract}

\maketitle

\section{Introduction}

Within the last few years, nonsequential double ionization (NSDI) in
strong, linearly polarized laser fields has attracted a great deal
of attention, both experimentally and theoretically
\cite{nsdireview}. This interest has been triggered by the outcome
of experiments in which the momentum component parallel to the laser
field polarization could be resolved, either for the doubly charged
ion \cite{expe1ion}, or for both electrons \cite{expe1}. Indeed, the
observed features, namely two circular regions along the parallel
momenta $p_{1\parallel }=p_{2\parallel }$ peaked at $p_{1\parallel
}=p_{2\parallel }=\pm 2\sqrt{U_{p}}$, with $U_{p}$ the ponderomotive
energy, are a clear fingerprint of electron-electron correlation,
and can be explained by a simple, three-step rescattering mechanism
\cite{corkum}. Thereby, an electron leaves an atom through tunneling
ionization (the \textquotedblleft first step\textquotedblright ),
propagates in the continuum, being accelerated by the field (the
\textquotedblleft second step\textquotedblright ), and recollides
inelastically with its parent ion (the \textquotedblleft third
step\textquotedblright ). In this collision, it transfers part of
its kinetic energy to a second electron, which is then released.

From the theoretical point of view, there exist models, both classical \cite%
{class,slowdown,pulse} and quantum-mechanical \cite%
{abeckerrep,abecker,smatrix,lphys,pransdi1,pransdi2,pulseqm}, based on such
a mechanism, which qualitatively reproduce the above-mentioned features.
They leave, however, several open questions. A very intriguing fact is that,
for instance, a very good agreement with the experiments is obtained if the
interaction through which the first electron is dislodged is of contact
type, and if no Coulomb repulsion is taken into account in the final
electron states. This agreement worsens if this interaction is modeled in a
more refined way, considering either a more realistic, Coulomb-type
interaction, or final-state electron-electron repulsion. Specifically, in
recent publications, such effects have been investigated in detail using
both an S-Matrix computation and a classical ensemble model, and have been
interpreted in terms of phase-space and dynamical effects \cite%
{pransdi1,pransdi2}. This analysis has been performed within the
Strong- Field Approximation (SFA) \cite{kfr}, which mainly consists
in neglecting the atomic binding potential in the propagation of the
electron in the continuum, the laser field when the electron is
bound or at the rescattering, and the internal structure of the atom
in question.

Within this framework, the NSDI transition amplitude is written as a
five-dimensional integral, with a time-dependent action and
comparatively slowly varying prefactors. Such an integral is then
solved using saddle-point methods. Apart from being less demanding
than evaluating such an integral numerically
\cite{abeckerrep,abecker}, or solving the time-dependent
Schr\"{o}dinger equation \cite{tdse}, these methods provide a clear
space-time picture of the physical process in question. In
particular, the results are interpreted in terms of the
so-called\textquotedblleft quantum orbits\textquotedblright. Such
orbits can be related to the orbits of classical electrons, and have
been extensively used in the context of above-threshold ionization,
high-order harmonic generation \cite{orbits} and, more recently,
nonsequential double ionization
\cite{lphys,pransdi1,pransdi2,pulseqm}.

The fact that, in \cite{pransdi1,pransdi2}, the crudest
approximation yields the best agreement with the experiments, seems
to indicate that the presence of the residual ion, which is not
taken into account, screens both the long-range interaction which
frees the second electron and the final-state repulsion. This
suggests that the presence of the ionic binding potential in the
physical steps describing nonsequential double ionization, i.e.,
tunneling, propagation and electron-impact ionization, should
somehow be incorporated. Indeed, in recent studies, it was found
that Coulomb focusing considerably influences the NSDI yield
\cite{coulombfocusing}.

 Another possibility is related to how the initial states in which the
 electrons are bound affect the electron-momentum distributions.
 Indeed, the poor agreement between the computations with the Coulomb
 interaction and the
experiments may be related to the fact that \emph{1s} states have
been used in this case, instead of states with a different shape or
spatial symmetry, such as, for instance, \emph{p} states.
Furthermore, it may as well be that an additional approximation
performed in \cite{pransdi1,pransdi2} for the contact interaction,
namely to assume that it takes place at the origin of the coordinate
system, contributes to the good agreement between theory and
experiments in this case. Physically, this means that the spatial
extension of the wave function of the second electron is neglected.
Such an approximation has not been performed in the computations for
the Coulomb interaction discussed in \cite{pransdi1,pransdi2}, and,
up to the present date, there exist no systematic studies of its
influence in the context of NSDI.

In this paper, we investigate such effects in the simplest possible
ways. First, we assume that both electrons are initially in
hydrogenic $2p$
and in $3p$ states, instead of in $s$ states, as previously done \cite%
{lphys,pransdi1,pransdi2,pulseqm}. One should note that, in contrast
to Helium, for which $s$ states are more appropriate, $p$ states
yield a more realistic description of the outer-shell electrons in
neon and argon, respectively. Since the two latter species are used
in most experiments, the choice of $p$ states is justified. This is
included in the transition amplitude as a form factor, and does not
modify the saddle-point equations. In both $p$ and $s$ - state
cases, we consider that the bound-state wave function of the second
electron is either localized at the origin or extends over a finite
spatial range, for the contact and Coulomb interactions. This
provides information on how the initial state of the \emph{second}
electron influences electron-impact ionization, and hence the NSDI
yield.

A further improvement consists in overcoming the bound-state singularity,
which is present in the saddle-point framework, and which has not been
addressed in \cite{pransdi1,pransdi2}. For this purpose, we use a slightly
modified action, with respect to that considered in \cite{pransdi1,pransdi2}%
, so that the tunneling process and the propagation of both
electrons in the continuum is altered. Such corrections depend on
the initial wave function of the \emph{first} electron. Hence, they
shed some light on how this wave function affects the
electron-momentum distributions. In particular, we investigate how
such corrections influence several features in the momentum
distributions, such as their shapes, the cutoff energies or the
contributions from different types of orbits to the yield.

Finally, we employ a modified form factor for the first electron,
upon return, which takes into account the ionic potential. This is a
first step towards incorporating the residual ion in our formalism.
As it will be discussed subsequently, this provides a strong hint
that the ion is important, in order to achieve a good agreement
between theory and experiment.

The manuscript is organized as following. In Sec. \ref{backgd}, we
provide the necessary theoretical background for understanding the
subsequent discussions. In Secs. \ref{pstates}, \ref{coulsing}, and
\ref{ion}, we present our results, and, finally, in Sec. \ref{concl}
we state our conclusions.

\section{Background}

\label{backgd}

\subsection{Transition amplitude}

\label{transampl}

The transition amplitude of the laser-assisted inelastic
rescattering process responsible for NSDI, in the strong-field
approximation and in atomic units, is given by
\begin{equation}
M=-\int_{-\infty }^{\infty }dt\int_{-\infty }^{t}dt^{\prime }\int
d^{3}kV_{\mathbf{p}_{j}\mathbf{,k}}V_{\mathbf{k,}0}\exp [iS(t,t^{\prime },%
\mathbf{p}_{j},\mathbf{k})],  \label{presc}
\end{equation}
with the action
\begin{eqnarray}
S(t,t^{\prime },\mathbf{p}_{j},\mathbf{k}) &=&-\frac{1}{2}%
\sum_{j=1}^{2}\int_{t}^{\infty }[\mathbf{p}_{j}+\mathbf{A}(\tau
)]^{2}d\tau
\label{action} \\
&&-\frac{1}{2}\int_{t^{\prime }}^{t}[\mathbf{k}+\mathbf{A}(\tau
)]^{2}d\tau +|E_{01}|t^{\prime }+|E_{02}|t. \notag
\end{eqnarray}
Eq.~(\ref{presc}) describes the following physical process: at a
time $t^{\prime},$ both electrons are bound ($|E_{01}|$ and
$|E_{02}|$ denote the first and second ionization potentials,
respectively). Then, the first electron leaves the atom by tunneling
ionization, and propagates in the continuum from the time
$t^{\prime}$ to the time $t$, only under the influence of the
external laser field $\mathbf{E}(t)=-d\mathbf{A}(t)/dt$. At this
latter time, it returns to its parent with intermediate momentum
$\mathbf{k}$, and gives part of its kinetic energy to the second
electron through the interaction $V_{12}$, so that it is able to
overcome the second ionization potential $|E_{02}|$. Finally, both
electrons reach the detector with final momenta $\mathbf{p}_{j}$
$(j=1,2)$. All the influence of the binding potential $V$ and of the
electron-electron interaction $V_{12}$ is included in the form
factors
\begin{equation}
V_{\mathbf{p}_{j},\mathbf{k}}=<\mathbf{p}_{2}+\mathbf{A}(t),\mathbf{p}_{1}+%
\mathbf{A}(t)|V_{12}|\mathbf{k}+\mathbf{A}(t),\psi _{0}^{(2)}>
\end{equation}
and \
\begin{equation}
V_{\mathbf{k,}0}=<\mathbf{k}+\mathbf{A}(t^{\prime })|V|\psi _{0}^{(1)}>,
\end{equation}%
which are explicitly given by
\begin{equation}
V_{\mathbf{k,}0}=\frac{1}{(2\pi )^{3/2}}\int \!\!d^{3}r_{1}\exp [i(\mathbf{k}%
+\mathbf{A(}t^{\prime }\mathbf{)}).\mathbf{r}_{1}]V(\mathbf{r}_{1})\psi
_{0}^{(1)}(\mathbf{r}_{1})
\end{equation}%
and
\begin{eqnarray}
V_{\mathbf{p}_{j},\mathbf{k}} &=&\frac{1}{(2\pi )^{9/2}}\int \int
d^{3}r_{1}d^{3}r_{2}\exp
[i(\mathbf{p}_{1}-\mathbf{k}).\mathbf{r}_{1}]
\notag \\
&&\!\!\!\times \exp [i(\mathbf{p}_{2}+\mathbf{A(}t\mathbf{)}).\mathbf{r}%
_{2}]V_{12}(\mathbf{r}_{2},\mathbf{r}_{1})\psi _{0}^{(2)}(\mathbf{r}_{2}),
\end{eqnarray}%
respectively. The binding potential $V(\mathbf{r}_{1})$ will be taken to be
of Coulomb type, and the interaction $V_{12}(\mathbf{r}_{2},\mathbf{r}_{1})$
through which the second electron is released will be chosen to be of
contact or Coulomb type. The initial state $\psi _{0}^{(1)}(\mathbf{r}_{1})$
of the first electron at the moment of its ionization will be taken as a
hydrogenic $s$ or $p$ state, and the wave function $\psi _{0}^{(2)}(\mathbf{r%
}_{2})$ of the second electron at the moment of its release is either chosen
as a hydrogenic state, or a Dirac delta state localized at $\mathbf{r}_{2}=0$%
. In Eq. (\ref{presc}), we neglect final-state electron-electron
repulsion (for a discussion of this effect see Refs.
\cite{abeckerrep,pransdi2}).

\subsection{Saddle-point analysis}
We solve Eq.~(\ref{presc}) applying the steepest descent method,
which is a very good approximation for low enough frequencies and
high enough driving-field intensities.  In this case, we must find
$\mathbf{k}$, $t^{\prime }$ and $t$ so that $S(t,t^{\prime
},\mathbf{p}_{n},\mathbf{k})$ $(n=1,2)$ is stationary, i.e., its
partial derivatives with respect to these parameters vanish. This
yields
\begin{equation}
\left[ \mathbf{k}+\mathbf{A}(t^{\prime })\right] ^{2}=-2|E_{01}|
\label{saddle1}
\end{equation}

\begin{equation}
\sum_{j=1}^{2}\left[ \mathbf{p}_{j}+\mathbf{A}(t)\right] ^{2}=\left[ \mathbf{%
k}+\mathbf{A}(t)\right] ^{2}-2|E_{02}|  \label{saddle2}
\end{equation}

\begin{equation}
\int_{t^{\prime }}^{t}d\tau \left[ \mathbf{k}+\mathbf{A}(\tau )\right] =0.
\label{saddle3}
\end{equation}%
Eq.~(\ref{saddle1}) gives the energy conservation during tunneling
ionization, and, for a non-vanishing ionization potential, has no
real solution. Consequently, $t,t^{\prime }$ and $\mathbf{k}$ are
complex quantities. In the limit $|E_{01}|\rightarrow 0,$
Eq.~(\ref{saddle1}) describes a classical electron leaving the
origin of the coordinate system with vanishing drift velocity.
Eq.~(\ref{saddle2}) expresses energy conservation at $t,$ in an
inelastic rescattering process in which the first electron gives
part of its kinetic energy to the second electron, so that it can
overcome the second ionization potential and reach the continuum.
Finally, Eq.~(\ref{saddle3}) yields the intermediate electron
momentum constrained by the condition that the first electron
returns to the site of its release.

The saddles determined by Eqs.~(\ref{saddle1})-(\ref{saddle3})
always occur in pairs that nearly coalesce at the boundaries of the
energy region for which electron-impact ionization is allowed to
occur, within a classical framework. Such a boundary causes the
yield to decay exponentially, leading to sharp cutoffs in the
momentum distributions.

If written in terms of the momentum components parallel and
perpendicular to the laser field polarization, Eq.~(\ref{saddle2})
reads
\begin{equation}
\sum_{j=1}^{2}\left[ p_{j||}+A(t)\right] ^{2}=\left[ \mathbf{k}+\mathbf{A}(t)%
\right] ^{2}-2|E_{02}|-\sum_{j=1}^{2}\mathbf{p}_{j\perp }^{2}
\label{circlep}
\end{equation}
and describes a hypersphere in the six-dimensional $(p_{j\parallel
}, \mathbf{p}_{j\perp })$ space.  This hypersphere delimits a region
in momentum space for which electron-impact ionization is
``classically allowed", i.e., exhibits a classical counterpart.
For constant transverse momenta, Eq. ~(\ref%
{circlep}) corresponds to a circle in the $(p_{1||},p_{2||})$ plane centered
at $-A(t)$ and whose radius is given by the difference between the kinetic
energy $E_{\mathrm{kin}}(t)=1/2\left[ \mathbf{k}+\mathbf{A}(t)\right] ^{2}$
of the first electron upon return and the effective ionization potential $|%
\tilde{E}_{02}|=|E_{02}|+\sum_{j=1}^{2}\mathbf{p}_{j\perp }^{2}/2.$
Clearly, this radius is most extensive if the final transverse
momenta $\mathbf{p}_{j\perp }),$ $(j=1,2)$ are vanishing, such as in
the examples provided in Sec. IV.

 In order to
compute the transition probabilities, we employ a specific uniform
saddle-point approximation, whose only applicability requirement is
that the saddles occur in pairs \cite{Bleistein,atiuni}. Unless
stated otherwise (e.g., in Sec. IV), we reduce the problem to two
dimensions, using Eq. (\ref{saddle3}) and the fact that the action
(\ref{action}) is quadratic in $\mathbf{k}$. Details about this
method, in the context of NSDI, are given in
\cite{lphys,pransdi1,pransdi2,pulseqm} (for above-threshold
ionization and high-order harmonic generation, c.f., \cite{atiuni}
and \cite{uniformhhg}, respectively).

The momentum distributions of electrons for various types of interaction $%
V_{12}$ read
\begin{equation}
M=\int d^{2}p_{1\bot }\int d^{2}p_{2\bot }|M_{L}+M_{R}|^{2}  \label{yield}
\end{equation}%
where the transverse momenta have been integrated over, and $M_{L}$
and $M_{R}$ give the left and the right peak in the momentum
distributions, respectively, computed using the uniform
approximation. We consider a monochromatic, linearly polarized
field, so that the vector potential reads
\begin{equation}
\mathbf{A}(t)=-A_{0}\cos (\omega t)\mathbf{e}_{x}.
\end{equation}%
In this case, $M_{R}=M(t,t^{\prime },p)$ and $M_{L}=M(t-T/2,t^{\prime
}-T/2,p),$ where $T=2\pi /\omega $ denotes a period of the driving field. We
use the symmetry property $|M(t,t^{\prime },p)|=|M(t-T/2,t^{\prime
}-T/2,-p)| $ to compute the left peak. One should\ note that, for other
types of driving fields, such as few-cycle pulses, this condition does not
hold and each peak must be computed independently \cite{pulse,pulseqm}.

\section{Initial p states}

\label{pstates}

Within the formalism discussed in the previous section, the first
and second electron, so far, have been assumed to be initially in
$1s$- or zero-range-potential bound states, whose energies
$|E_{01}|$ and $|E_{02}|$ are taken to be the first and second
atomic ionization potential, respectively. In most experiments,
however, species such as neon and argon are used, for which the
outer-shell electrons are in $2p$ and $3p$ states, respectively. For
this reason, such states should provide a more realistic modeling of
laser-induced nonsequential double ionization. For symmetry reasons,
only the states with magnetic quantum number $m=0$ will contribute
to the yield.

In this case, the bound-state wave functions of both electrons will be given
by
\begin{equation}
\psi _{210}^{(j)}(r_{j})=C_{210}r_{j}\exp [-\alpha_j r_{j}]\cos
\theta \label{2p}
\end{equation}%
and\
\begin{equation}
\psi _{310}^{(j)}(r_{j})=C_{310}r_{j}(1-\alpha_j r_{j}\ /2)\exp
[-\alpha_j r_{j}]\cos \theta ,  \label{3p}
\end{equation}%
respectively, where $\alpha_j =\sqrt{2|E_{0j}|}(j=1,2)$, and
$C_{n10}$ (where $n$ is the principal quantum number) denote
normalization constants. For comparison, we will also consider hydrogenic $%
1s $ wave functions, which read
\begin{equation}
\psi _{100}^{\ (j)}(r_{j})=C_{100}\exp [-\alpha_j r_{j}]. \label{1s}
\end{equation}
In Eqs. (\ref{2p})-(\ref{1s}),the binding energies of the first and
the second electron were chosen as the first and the second
ionization potentials, respectively.

The form factors $V_{\mathbf{p}_{j}\mathbf{,k}},$ for $2p$ and $3p$
initial states, read
\begin{equation}
V^{(2p)}_{\mathbf{p}_{j}\mathbf{,k}}\sim \eta(\mathbf{p}_1,\mathbf{k})\frac{\tilde{p}}{\left[ 2|E_{02}|+%
\mathbf{\tilde{p}}^{2}\right] ^{3}}+(\mathbf{p}_{1}\leftrightarrow \mathbf{p}%
_{2})  \label{delt2p}
\end{equation}%
and
\begin{equation}
V^{(3p)}_{\mathbf{p}_{j}\mathbf{,k}}\sim \eta(\mathbf{p}_1,\mathbf{k})\frac{\tilde{p}(\mathbf{\tilde{p}}%
^{2}-2|E_{02}|)}{\left[ 2|E_{02}|+\mathbf{\tilde{p}}^{2}\right]
^{4}}+(\mathbf{p}_{1}\leftrightarrow \mathbf{p}%
_{2}), \label{delt3p}
\end{equation}%
respectively, with $\mathbf{\tilde{p}=p}_{1}+\mathbf{p}_{2}-\mathbf{k}+%
\mathbf{A}(t)$. Thereby, $(\mathbf{p}_{1}\leftrightarrow \mathbf{p}%
_{2})$ means that the momenta of both particles are interchanged, and
$\eta(\mathbf{p}_j,\mathbf{k})$ $(j=1,2)$ is a function which depends on the interaction in question. The corresponding form factor obtained for an initial state (%
\ref{1s}) is given by
\begin{equation}
V^{(1s)}_{\mathbf{p}_{j}\mathbf{,k}}\sim \eta(\mathbf{p}_1,\mathbf{k})\frac{1}{\left[ 2|E_{02}|+\mathbf{%
\tilde{p}}^{2}\right] ^{2}}+(\mathbf{p}_{1}\leftrightarrow \mathbf{p}%
_{2}).  \label{deltshortr}
\end{equation}

\subsection{Contact-type interaction}

\label{pstcontact}

As a first step, we will assume that the second electron is released by a
contact-type interaction
\begin{equation}
V_{12}\sim\delta (\mathbf{r}_{1}-\mathbf{r}_{2}).  \label{contact}
\end{equation}%
In this case, in Eqs. (\ref{delt2p})-(\ref{deltshortr}),
$\eta(\mathbf{p}_j,\mathbf{k})$ is a constant. The differential
electron momentum distributions computed with such form
factors are depicted in Figs. 1(a)-1(c), as contour plots in the $%
(p_{1||},p_{2||})$ plane. In such computations, only the pair of orbits for
which the electron excursion times $\tau =t-t^{\prime }$ in the continuum
are shortest have been employed. As an overall feature, the distributions
are peaked near $p_{1||}=p_{2||}=\pm \sqrt{U_{p}}$ and spread in the
direction perpendicular to the diagonal $p_{1||}=p_{2||}$.
\begin{figure}[tbp]
\includegraphics[width=8.5cm]{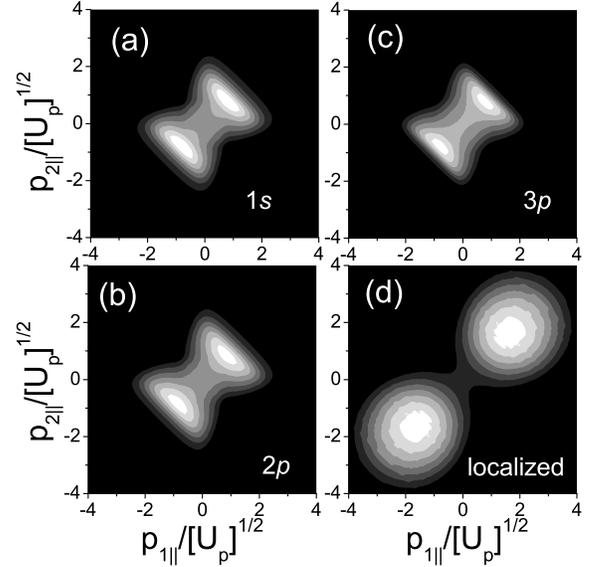}
\caption{Electron momentum distributions computed using a contact-type
interaction, as functions of the electron momentum components parallel to
the laser-field polarization. The external field was taken to be
monochromatic, with frequency $\protect\omega =0.057$ a.u. and intensity $%
I=8\times 10^{14}\mathrm{W/cm}^{2}$ ($U_p=1.75$ a.u.). In panels
(a), (b) and (c), both electrons are initially bound in a $1s$,
$2p$, and $3p$ state [Eqs. (\protect\ref{1s}), (\protect\ref{2p})
and (\protect\ref{3p})], respectively, whereas, in part (d), the
first electron is initially in a $1s$ state and the wave function of
the second electron is localized at $r_{2}=0$. In all situations
(even for the $3p$ - state case), the atomic species was taken to be
neon ($|E_{01}|=0.79$ a.u. and $|E_{02}|=1.51$ a.u.), in order to
facilitate a clear assessment of the effects caused by the different
initial states. The transverse momenta have been integrated over.}
\end{figure}

An inspection of the form factors (\ref{delt2p})-(\ref{deltshortr}),
for constant $\eta(\mathbf{p}_j,\mathbf{k})$, explains this
behavior. Indeed, such form factors are large if their denominator
is small. Since
 $|E_{02}|$ is constant, this condition implies that $\mathbf{\tilde{p}=p}%
_{1}+\mathbf{p}_{2}-\mathbf{k}+\mathbf{A}(t)$ is small. To first
approximation, since the first electron returns at times $t$ close
to the minimum of the electric field, one may assume that the vector
potential at this time and the intermediate electron momentum are
approximately constant. Furthermore, in the model, the field is
approximated by a monochromatic wave and\textbf{\ }$\mathbf{k}$ is
given by Eq. (\ref{saddle3}). Hence, a rough
estimate of these quantities at the return times yields $A(t)\simeq 2\sqrt{%
U_{p}}$ and $k\simeq 0$, respectively. Thus, $\mathbf{\tilde{p}}$ will be
small mainly if $\mathbf{p}_{1}=-\mathbf{p}_{2}$, so that contributions
along the anti-diagonal $p_{1||}=-p_{2||}$ will be enhanced.

Such contributions get more localized near the maxima for highly excited
initial states due to the increase in the exponent of the denominator. A
direct look at the above-stated form factors confirms this interpretation,
yielding maxima along the anti-diagonal and near $\pm \sqrt{U_{p}}$.

Interestingly, the distributions obtained for the contact interaction are
quite different from the circular distributions peaked around $%
p_{1||}=p_{2||}=\pm 2\sqrt{U_{p}}$ observed experimentally. Indeed, in order
to obtain such distributions, it is necessary to assume that the initial
wave function of the second electron is localized at $r_{2}=0.$ This is
formally equivalent to taking
\begin{equation}
V_{12}(\mathbf{r}_{1}-\mathbf{r}_{2})\psi _{0}^{(2)}(\mathbf{r}%
_{2})\sim\delta (\mathbf{r}_{1}-\mathbf{r}_{2})\delta (\mathbf{r}_{2}).
\label{contorigin}
\end{equation}
Eq. (\ref{contorigin}) yields a constant form factor $V_{\mathbf{p}_j,%
\mathbf{k}}$. In Fig. 1.(d), we present the distributions computed using Eq.
(\ref{contorigin}), which exhibit a very good agreement with the
experiments. This means that, in reality, the effective wave function of the
second electron is very localized, most probably due to refocussing \cite%
{coulombfocusing}, or screening effects \cite{footnscreening}.

\subsection{Coulomb-type interaction}

\label{pstcoul}

We will now consider that the second electron is released by a Coulomb type
interaction, given by
\begin{equation}
V_{12}=1/|\mathbf{r}_{1}-\mathbf{r}_{2}|.  \label{Coulomb}
\end{equation}%
In this case, in the form factors (\ref{delt2p})-(\ref{deltshortr}),
$\eta$ is given by
\begin{equation}
\eta(\mathbf{p}_j,\mathbf{k})=\frac{1}{[\mathbf{p}_{j}-\mathbf{k%
}]^{2}} \hspace*{0.5cm}(j=1,2),\label{etacoul}
\end{equation}
respectively. This causes the prefactors to be large
 when $\mathbf{p}_{j}-%
\mathbf{k}$ $(j=1,2)$ is small, in addition to the case for which
$\mathbf{\tilde{p}}\ll 1$. The influence of such form factors on the
electron momentum distributions is shown in Fig. 2. Apart from the
broadening along $p_{1||}=-p_{2||}$ caused
by the spatial extent of the bound-state wave functions (c.f. Sec. \ref%
{pstcontact}), the distributions exhibit maxima near the axis
$p_{1||}=0$ or $p_{2||}=0.$ Such maxima are due to the factor
(\ref{etacoul}) in Eqs. (\ref{delt2p})-(\ref{deltshortr}),
characteristic of the Coulomb-type interaction, which is large for
$\mathbf{p}_{j}\simeq \mathbf{k} $. Since, to first approximation,
contributions from regions of small $k$
dominate the yield, one expects maxima in momentum regions where either $%
\mathbf{p}_{1}$ or $\mathbf{p}_{2}$ are small.

Furthermore, as compared to the yields obtained using a Coulomb-type
interaction and $1s$-states, there exists a small additional
broadening in the distributions, with respect to the diagonal
$p_{1||}=p_{2||}$, as well as an increase in the contributions from
regions where such momenta are small. Such effects get more
pronounced as the principal quantum number increases, as shown in
Figs. 2.(b) and 2.(c).

However, such modifications do not alter the distributions in a
significant way. More extreme changes occur, for instance, if a
contact-type interaction, i.e.,
$\eta(\mathbf{p}_j,\mathbf{k})=const.$ in Eq.~(\ref{contact}), is
taken into account. Still, less localized bound states for both
electrons will cause a broadening in the momentum distributions. If
the second-electron wave function is localized at the origin, the
form factor (\ref{deltshortr}) reduces to
\begin{equation}
V_{\mathbf{p}_{j}\mathbf{,k}}\sim \frac{1}{[\mathbf{p}_{1}-\mathbf{k%
}]^{2}}+(\mathbf{p}_{1}\leftrightarrow \mathbf{p}_{2}).
\label{coullim}
\end{equation}%
The distributions for the latter form factor are displayed in Fig. 2.(d). In
the figure, one observes a considerable reduction of the broadening along
the anti-diagonal $p_{1||}=-p_{2||}$. However, the distributions still
exhibit the two sets of maxima near the axis $p_{1||}=0$ or $p_{2||}=0.$
This is expected, since such maxima are a fingerprint of the Coulomb
interaction.
\begin{figure}[tbp]
\includegraphics[width=8.5cm]{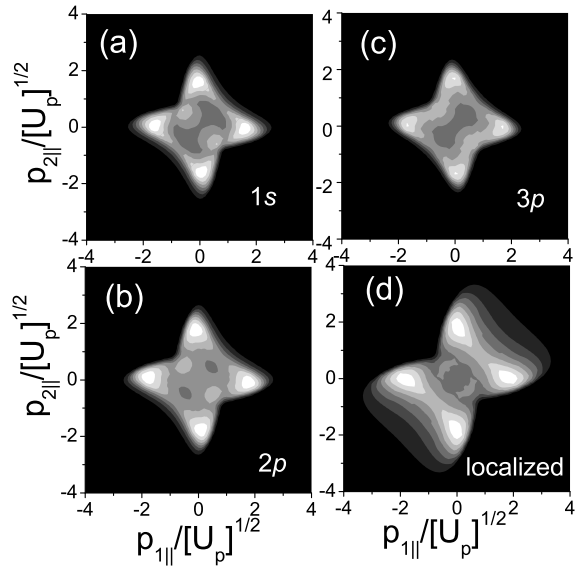}
\caption{Electron momentum distributions computed using a Coulomb-type
interaction, as functions of the electron momentum components parallel to
the laser-field polarization. The field and atomic parameters are the same
as in Fig. 1. In panels (a), (b) and (c), both electrons are taken to be
initially in a $1s$, $2p$, and $3p$ state, respectively, whereas in panel
(d) the first electron is in a $1s$ state, while the spatial extension of
the bound-state wave function has been neglected. The transverse momenta
have been integrated over.}
\end{figure}

The results in this section show that the shapes of the momentum
distributions in NSDI are not only influenced by the type of
interaction by which the second electron is dislodged but,
additionally, depend on the spatial extension of the wave function
of the state where it is initially bound. In fact, radically
different shapes are observed if this wave function is either taken
to be localized at $\mathbf{r}_{2}=0$ or exponentially decaying.
This is true both for a contact- and a Coulomb-type interaction
(c.f. Figs. 1.(d) and 2.(d)).

On the other hand, if different initial states are taken, for the same type
of interaction, there are no significant changes in the shapes of the
distributions as long as such states extend over a finite spatial range.
This is explicitly seen by comparing yields obtained using bound states with
different principal quantum numbers. This is related to the fact that the
wave functions (\ref{2p})-(\ref{1s}) were chosen such that the bound-state
energy always corresponds to the second ionization potential. Hence, even if
their shape changes, the spatial extension of such wave functions is roughly
the same.

It is still, however, quite puzzling that the best agreement with the
experimental findings occurs for the crudest approximations, both for the
interaction and the initial bound-state wave function, i.e., for a
contact-type interaction and a wave function localized at $\mathbf{r}_{2}=0.$
Indeed, taking either a more realistic type of electron-electron
interaction, spatially extended bound states, or, still, bound states which
are, in principle, a more refined description of the outer-shell electrons,
only worsens the agreement between experiment and theory.

If the main physical mechanism of NSDI is electron-impact
ionization, there exist two main possibilities for explaining this
discrepancy. Either the second electron is bound in a highly
localized state and both electrons collide through an effective
short-range interaction, as the present results suggest, or the
tunneling ionization, as well as the electron propagation in the
continuum, must be improved. The first issue may be addressed by
including the influence of the residual ion in the process, whereas
the second issue may be dealt with in several ways. For instance, in
the subsequent section, we will consider corrections of a more
fundamental nature, which alter the semiclassical action and thus
the orbits of the electrons.

\section{Treatment of the bound-state singularity}

\label{coulsing}

Up to the present section, we have implicitly assumed that the form factors $%
V_{\mathbf{p}_{{j}}\mathbf{,k}}$ and $V_{%
\mathbf{k,}0}$ are free of singularities and slowly varying in comparison to
the time-dependent action. However, this is not always true. Indeed, in the
saddle-point framework, the form factor $V_{\mathbf{k,}0}$ is singular if
the electron is initially in a state described by an exponentially decaying
wave function, such as Eqs. (\ref{2p})-(\ref{1s}). More specifically, in
this case,%
\begin{equation}
V_{\mathbf{k,}0}\varpropto \frac{f(k+A(t^{\prime }))}{\left( \left[ \mathbf{k%
}+\mathbf{A}(t^{\prime })\right] ^{2}+2|E_{01}|\right) ^{n}},
\end{equation}%
where $n$ is an integer number. In this case, according to Eq.
(\ref{saddle1}), the denominator vanishes. Due to this singularity,
this form factor does not vary slowly with respect to the
semi-classical action (\ref{action}), and thus must be incorporated
in the
exponent. Therefore, we take the modified action%
\begin{equation}
\tilde{S}(t,t^{\prime },\mathbf{p}_{j},\mathbf{k})=S(t,t^{\prime },\mathbf{p}%
_{j},\mathbf{k})-i\ln \left[ V_{\mathbf{k,}0}\right]  \label{smod}
\end{equation}%
in the transition amplitude (\ref{presc}). This causes a change in the first
and third saddle-point equations, which will depend on the initial bound
state in question. In particular, we will consider that the first electron
is initially in the hydrogenic states $1s$, $2p$, and $3p$. This is a
legitimate assumption, since the binding potential of a neutral atom, from
which the first electron tunnels out, is of long-range type. For the states $%
1s$, $2p$, and $3p$, $V_{\mathbf{k},0}$ reads%
\begin{equation}
V_{\mathbf{k},0}^{(1s)}=\frac{\sqrt{2}}{\pi }\frac{(2|E_{01}|)^{5/2}}{%
\mathbf{v}^{2}+2|E_{01}|},  \label{vk01s}
\end{equation}

\begin{equation}
V_{\mathbf{k},0}^{(2p)}=\frac{2\sqrt{2}i}{\pi }\frac{(\sqrt{2|E_{01}|}%
)^{5/2}v}{(\mathbf{v}^{2}+2|E_{01}|)^{2}},  \label{vk02p}
\end{equation}%
and%
\begin{equation}
V_{\mathbf{k,}0}^{(3p)}=\frac{8i(\sqrt{2|E_{01}|})^{5/2}v(\mathbf{v}%
^{2}-2|E_{01}|)}{\sqrt{3}\pi (\mathbf{v}^{2}+2|E_{01}|)^{3}},  \label{vk03p}
\end{equation}%
respectively, where $\mathbf{v}=\left[ \mathbf{k}+\mathbf{A}(t^{\prime })%
\right] $ denotes the initial electron drift velocity. The explicit
expressions for the saddle point equations then become
\begin{equation}
\left[ \mathbf{k}+\mathbf{A}(t^{\prime })\right] ^{2}=-2|E_{01}|+\mathbf{%
\zeta }(k,t^{\prime }).\mathbf{E}(t^{\prime })  \label{sadmod1}
\end{equation}%
and%
\begin{equation}
\int_{t^{\prime }}^{t}d\tau \left[ \mathbf{k}+\mathbf{A}(\tau )\right] +%
\mathbf{\zeta}(k,t^{\prime })=0,  \label{sadmod3}
\end{equation}%
respectively, where $\mathbf{\zeta }(k,t^{\prime })=-i\partial _{\mathbf{k}%
}\ln [V_{\mathbf{k,}0}]$ is a correction which depends on the
initial bound state. Thus, there is an effective shift in the
ionization potential at the tunneling times, and a modification in
the return condition. Consequently, the orbits change. Apart from
that, from the technical point of view, the transition amplitude is
no longer reducible to a two-dimensional integral, so that the
problem is far more cumbersome.

The modifications in the equation describing tunneling ionization allow the
existence of solutions for which $\mathrm{Re}[v]\neq 0$. This did not occur
in Eq. (\ref{saddle1}), for which this quantity was purely imaginary, and,
physically, means that there are in principle changes, maybe even
enhancements, in the probability that the first electron tunnels out at $%
t^{\prime }.$

Furthermore, Eq. (\ref{sadmod3}), if written in terms of the components of
the intermediate momentum $\mathbf{k}$ parallel and perpendicular to the
laser field polarization, has, apart from the trivial solution $\mathbf{k}%
_{\perp }=0,$ additional solutions for which $\mathbf{k}_{\perp
}\neq 0.$ Thus, in principle, the first electron may have, during
the tunnel ionization and upon return, a non-vanishing drift
velocity component transverse to the laser-field polarization. We
regard this possibility, however, as non-physical, and therefore
will mainly concentrate on the case of vanishing $\mathbf{k}_{\perp
}.$ Despite of that, the results obtained for nonvanishing
$\mathbf{k}_{\perp }$will be briefly discussed in Sec.
\ref{kstrange}. For the return condition (\ref{saddle2}) this is not
possible and $\mathbf{k}_{\perp }$ is always vanishing. In the
following, we will investigate how the corrections in the action
affect the momentum distributions.

\subsection{Vanishing $\mathbf{k}_{\perp }$}

In this section, we will consider that the first electron has vanishing
intermediate momentum components $\mathbf{k}_{\perp }$. Physically, this
means that the dynamics of NSDI is mainly taking place along the laser field
polarization, which is the intuitively expected situation. In Fig. 3, we
present the electron momentum distributions computed employing the modified
saddle-point equations and the action (\ref{smod}), for the same initial
states and types of as in Figs. 1 and a contact-type interaction. In
general, the distributions in Fig. 3 are very similar to the former ones,
with, however, a suppression in the region of small parallel momenta. This
is true even if different corrections are taken into account, as it is the
case if the first electron is initially in a $1s$, $2p$ and $3p$ state
(Figs. 3.(a), 3.(b) and 3.(c), respectively). In the specific case of a
localized bound-state wave function for the second electron, there is also a
minor displacement of the maxima towards smaller parallel momenta(c.f. Fig.
3(d)).

The suppression persists if the second electron is released by a
Coulomb-type interaction, as shown in Fig. 4. Specifically for this
interaction, the corrections lead to a suppression of the secondary maxima
in the small-momentum region, which were present in Fig. 2.
\begin{figure}[tbp]
\includegraphics[width=8.5cm]{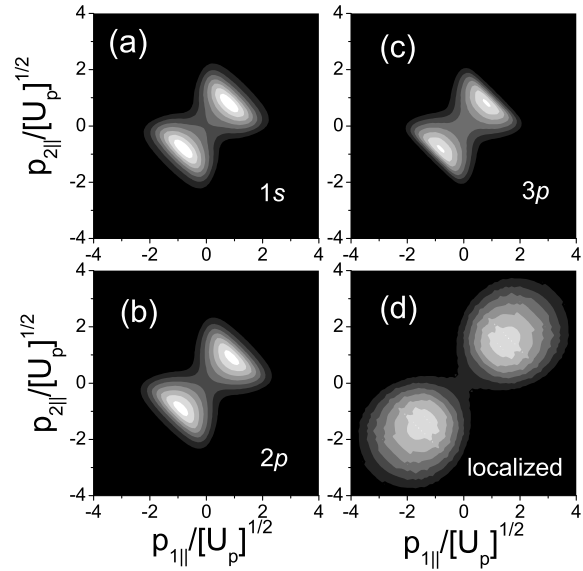}
\caption{Electron momentum distributions computed with using contact-type
interaction, as functions of the electron momentum components parallel to
the laser-field polarization, for the same field and atomic parameters as in
the previous figures. We introduce corrections to the bound-state
singularity by employing the modified action (\protect\ref{smod}) and
saddle-point equations, taking the solutions displayed in Figs. 6.(a) and
6.(c). In panels (a), (b) and (c), both electrons are taken to be initially
in a $1s$, in a $2p$, and in a $3p$ state, respectively, whereas in panel
(d) the first electron is initially in a 1s state, while the spatial
extension of the bound-state wave function of the second electron has been
neglected. The transverse momenta have been integrated over.}
\end{figure}

\begin{figure}[tbp]
\includegraphics[width=8.5cm]{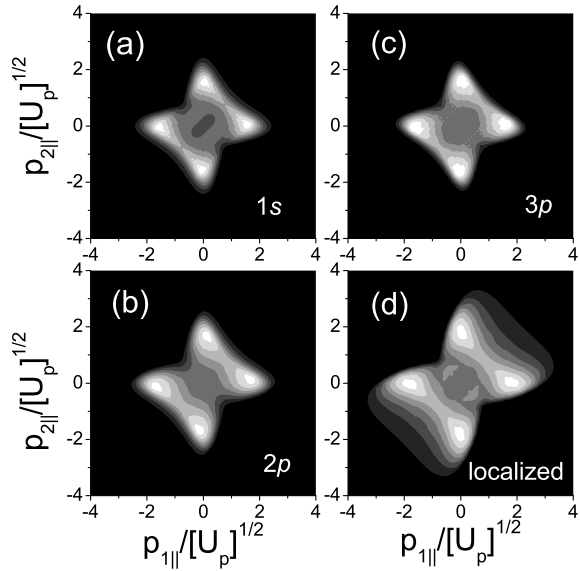}
\caption{Electron momentum distributions computed using the Coulomb-type
interaction, as functions of the electron momentum components parallel to
the laser-field polarization, for same field and atomic parameters as in the
previous figure. We introduce corrections to the bound-state singularity by
employing the modified action (\protect\ref{smod}) and saddle-point
equations, and the solutions in Figs. 6(a) and 6.(c). In panels (a), (b) and
(c), both electrons are taken to be initially in a $1s$, in a $2p$, and in a
$3p$ state, respectively, whereas in panel (d) the first electron is
initially in a 1s state, while the spatial extension of the bound-state wave
function of the second electron has been neglected. The transverse momenta
have been integrated over.}
\end{figure}

In the following, we will analyze these differences in terms of the
so-called quantum orbits, obtained by solving the saddle point equations. We
will consider both the saddle-point equations in the presence and absence of
corrections to the bound-state singularity, i.e., Eqs. (\ref{sadmod1}), (\ref%
{saddle2}) and (\ref{sadmod3}), and (\ref{saddle1})-(\ref{saddle3}),
respectively. We restrict ourselves to vanishing final transverse momenta
and longitudinal momentum components along the diagonal $%
p_{||}=p_{1||}=p_{2||}$. For this particular case, the energy region
for which electron-impact ionization is classically allowed is most
extensive.
\begin{figure}[tbp]
\includegraphics[width=8.5cm]{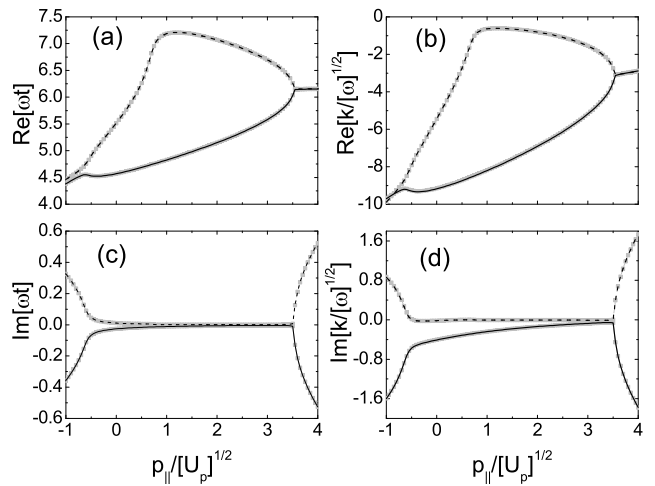}
\caption{Rescattering times, together with the intermediate momentum $k_{||}$%
, as functions of the parallel momentum $p_{||}$ along the diagonal $%
p_{1||}=p_{2||}$. The final transverse momenta $\mathbf{p}_{j\perp
}(j=1,2)$ and the intermediate transverse momentum
$\mathbf{k}_{\perp }$ are taken to be vanishing. The real and
imaginary parts of such quantities are displayed in the upper [(a),
(b) and (c)] and lower [(d), (e) and (f)] panels, respectively. The
field and atomic parameters are the same as in the previous figure.
\ The uncorrected variables are given by the thick light gray curves
in the figure, while the variables with corrections corresponding \
to initial $1s$, $2p$ and $3p$ states are given by the gray, dark
gray and black curves, respectively. The longer and the shorter
orbit are indicated by dashed and solid lines, respectively.}
\end{figure}
\begin{figure}[tbp]
\includegraphics[width=8.5cm]{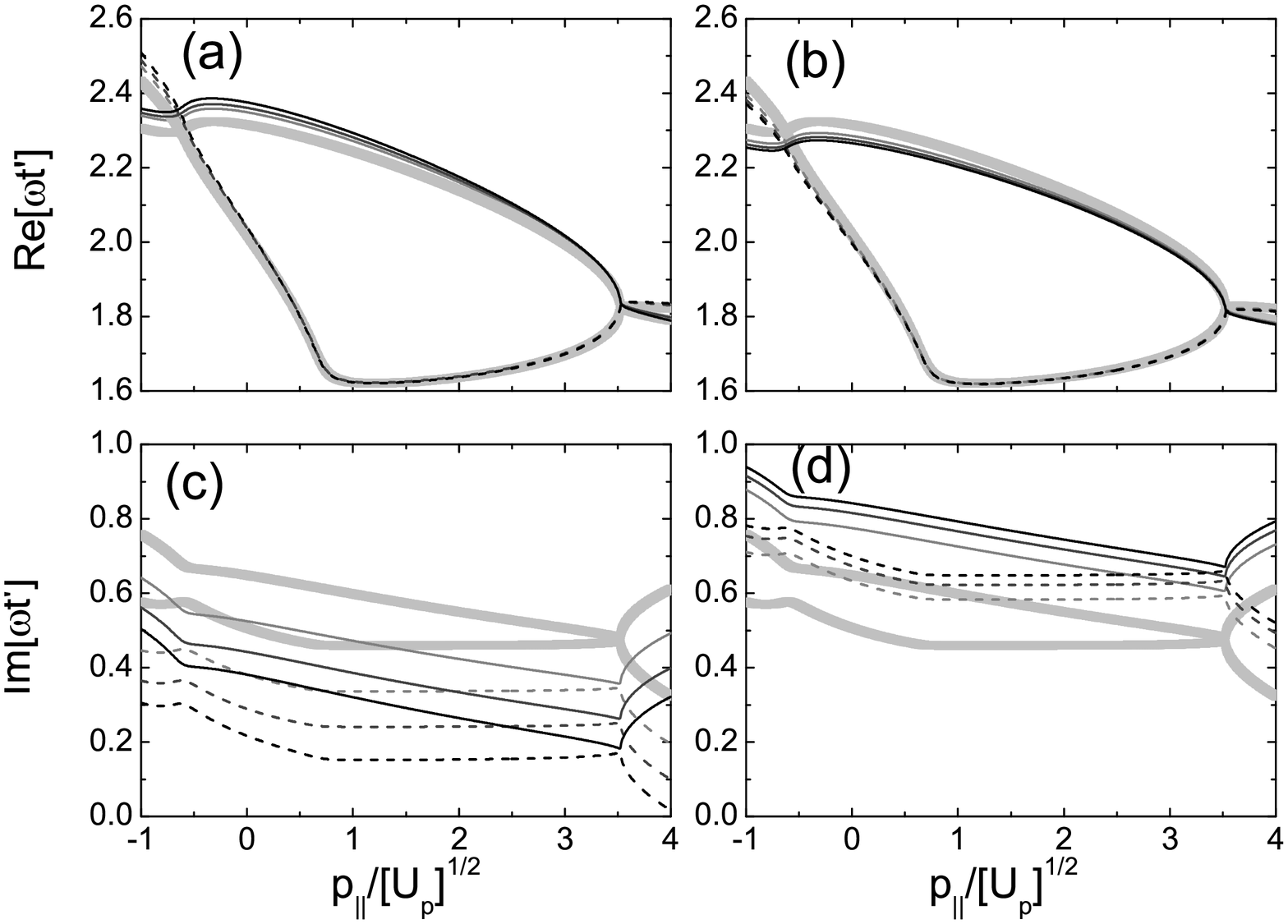}
\caption{Tunneling times as functions of the parallel momentum $p_{||}$
along the diagonal $p_{1||}=p_{2||}$. The final transverse momenta $\mathbf{p%
}_{j\perp }(j=1,2)$ and the intermediate transverse momentum $\mathbf{k}%
_{\perp }$ are taken to be vanishing. The real and imaginary parts of such
quantities are displayed in the upper [(a) and (b)] and lower [(c) and (d)]
panels, and the same pairs of orbits are depicted in the left [(a) and (c)]
and right [(b) and (d)] panels, respectively. The field and atomic
parameters are the same as in the previous figure. \ The uncorrected
variables are given by the thick light gray curves in the figure, while the
variables with corrections corresponding \ to initial $1s$, $2p$ and $3p$
states are given by the gray, dark gray and black curves, respectively. The
longer and the shorter orbit are indicated by dashed and solid lines,
respectively.}
\end{figure}

In Fig. 5, we display the solutions of the saddle-point equations
for the rescattering times $t$ and the intermediate momentum
$\mathbf{k}$. The upper and lower panels in the figure give the real
and imaginary parts of such variables, respectively. The real parts
of $t$ and $\mathbf{k}$ correspond to the solutions of the equations
of motion of a classical electron in an external laser field, and
almost merge at two distinct parallel momenta. These momenta are
related to the maximal and minimal energy for which the second
electron is able to overcome $|E_{02}|.$ Beyond such momenta, there
are cutoffs in the distributions, and the yield decays
exponentially. The imaginary parts of such variables are in a sense
a measure of a particular physical process being classically allowed
or forbidden. Indeed, the fact that $|\mathrm{Im}[t]|$ and
$|\mathrm{Im}[k]|$ are vanishingly small between the minimal and
maximal allowed momenta are a consequence of both electron-impact
ionization and the return condition being classically
allowed in this region. As the boundaries of this region are reached, $|%
\mathrm{Im}[t]|$ and $|\mathrm{Im}[k]|$ increase exponentially.
Interestingly, both the real and imaginary parts of such variables, as well
as the cutoff momenta, remain practically inaltered upon the changes
introduced in this section. This is not obvious, since the bound-state
corrections in question alter the return condition [c.f. Eq.(\ref{sadmod3})].

There exist, however, modifications in the tunneling times $t^{\prime }$,
which are explicitly shown in Fig. 6. Specifically, the corrections in the
tunneling condition, which leads to Eq. (\ref{sadmod1}), cause a splitting
in the solutions of Eq. (\ref{saddle1}). This follows from the fact that
small variations in the stationary-action trajectories contributes
quadratically to $S(t,t^{\prime},\mathbf{p}_j,\mathbf{k})$ and $V_{\mathbf{k}%
,0}$, so that $\tilde{S}(t,t^{\prime },\mathbf{p}_{j},\mathbf{k})$
attains two stationary trajectories for each of the former ones.
Strictly speaking, a similar splitting also occurs for $\mathbf{k}$
and $t$. In practice, however, the difference between the two
different sets of solutions is vanishingly small, and thus not
noticeable in Fig. 5. The different sets of solutions are depicted
in Figs. 6(a) and 6.(c), and 6.(b) and 6(d), respectively. \ The
real parts $\mathrm{Re}[t^{\prime }]$ exhibit only minor
differences, with occur for the shorter orbits and small momenta and
eventually disappear as the upper cutoff is approached. Depending on
the type of correction, such times either distance themselves from,
or become slightly closer to the peak-field times (Figs. 6.(a) and
6.(b), respectively). Thus, one could expect an enhancement in the
contributions from the shorter orbits near the origin of the $\left(
p_{1||},p_{2||}\right) $ plane, in the former case, and a
suppression in the latter case. However, we have used the solutions
in Fig. 6.(b) and (d) for computing the contour plots in Figs. 3 and
4, and obtained a suppression in
the yield. This is a clear indication that the changes in $\mathrm{Im}%
[t^{\prime }]$ and in the time-dependent action play a more important role
than those in $\mathrm{Re}[t^{\prime }]$.

In Figs. 6.(c) and 6.(d), we present the imaginary parts of $t^{\prime },$
which clearly shift towards smaller, and larger values, respectively, when
the corrections $\mathbf{\varsigma }(k,t^{\prime })$ are taken into account.
The higher the initial state lies, the larger such shifts are. Physically,
there exists a correspondence between such imaginary parts and the
probability that the first electron tunnels out and reaches the continuum.
This means that, by using a slightly modified action in order to overcome
the Coulomb singularity, one is changing the effective potential barrier at $%
t^{\prime }$ for the first electron. In general, such a barrier has a
significant influence on the distributions. Indeed, recently, we have shown,
within the context of nonsequential double ionization with few-cycle laser
pulses, that the importance of the contributions of a particular orbit or
set of orbits to the yield is highly dependent on $|\mathrm{Im}[t^{\prime
}]| $. The smaller this quantity is, the larger is the tunneling probability
for the first electron \cite{pulseqm}. As a direct consequence,
contributions from orbits with small $|\mathrm{Im}[t^{\prime }]|$, i.e.,
with a large tunneling probability, dominate the yield. In the present case,
however, since both orbits are being equally shifted, this should not
influence the distributions qualitatively. One should note that, even in the
momentum region for which electron-impact ionization is allowed, $|\mathrm{Im%
}[t^{\prime }]|$ is always nonvanishing. This is a direct consequence of the
fact that tunneling ionization is a classically forbidden process.

Subsequently, we compute the counterparts of Fig. 3 and 4 (Figs. 7 and 8)
using the solutions displayed in Fig. 6.(b) and 6.(d). Also in this case, in
general, there is a suppression in the yield in the region of small parallel
momenta, with, however, a slightly different substructure in the
Coulomb-interaction case.
\begin{figure}[tbp]
\includegraphics[width=8.5cm]{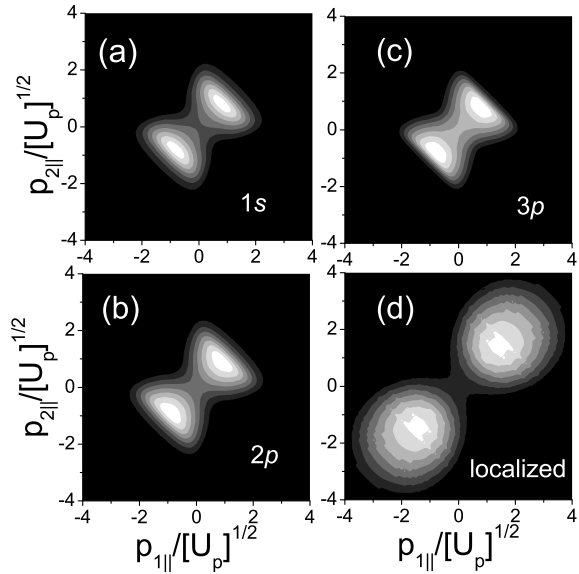}
\caption{Electron momentum distributions computed with the
contact-type interaction, as functions of the electron momentum
components parallel to the laser-field polarization, for the same
field and atomic parameters as in the previous figure. We introduce
corrections to the bound-state singularity by employing the modified
action (\protect\ref{smod}) and saddle-point equations, taking the
solutions in Figs. 6.(b) and 6.(d). In panels (a), (b) and (c), both
electrons are taken to be initially in a $1s$, in a $2p$, and in a
$3p$ state, respectively, whereas in panel (d) the first electron is
in a $1s$ state, and spatial extension of the bound-state wave
function of the second electron has been neglected. The transverse
momenta have been integrated over.}
\end{figure}
\begin{figure}[tbp]
\includegraphics[width=8.5cm]{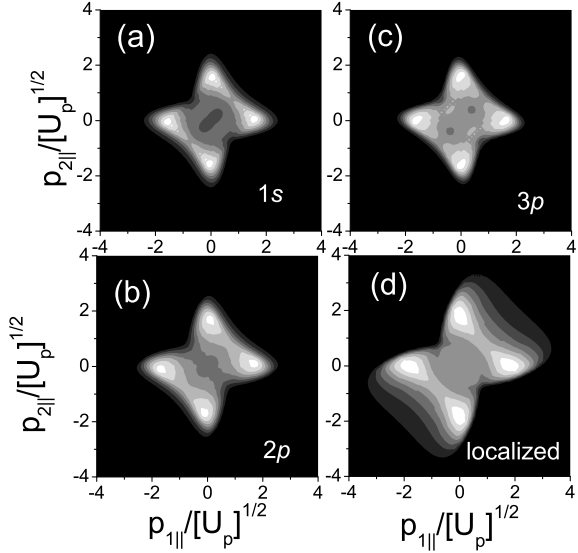}
\caption{Electron momentum distributions computed using the Coulomb-type
interaction, as functions of the electron momentum components parallel to
the laser-field polarization, for same field and atomic parameters as in the
previous figure. We introduce corrections to the bound-state singularity by
employing the modified action (\protect\ref{smod}) and saddle-point
equations, and the solutions in Figs. 6.(b) and 6.(d). In panels (a), (b)
and (c), both electrons are taken to be initially in a $1s$, in a $2p$, and
in a $3p$ state, respectively, whereas in panel (d) the first electron is
initially in a 1s state, while the spatial extension of the bound-state wave
function of the second electron has been neglected. The transverse momenta
have been integrated over.}
\end{figure}

\subsection{Nonvanishing $\mathbf{k}_{\perp }$}

\label{kstrange}

The modifications introduced in the return condition for the first electron
[Eq. (\ref{sadmod3})] allow the intermediate momentum $\mathbf{k}$ to have a
nonvanishing component perpendicular to the laser-field polarization. \ This
implies that the first electron, during tunneling ionization and when it
returns, is being deviated from its original direction. Although such an
effect is unphysical, we will briefly discuss its consequences. \ For that
purpose, we will consider the simplest corrections to the bound-state
singularity discussed in this paper, namely those for $1s$ initial states.
If Eq. (\ref{sadmod3}) is written in terms of the intermediate-momentum
components $\mathbf{k}_{\perp }$ and $k_{||}$ perpendicular and parallel to
the laser-field polarization, this equation reads%
\begin{equation}
k_{||}(t-t^{\prime })-\int_{t^{\prime }}^{t}A(s)ds+\frac{2i[k_{||}+A(t^{%
\prime })]}{\left[ 2|E_{01}|+\mathbf{k}_{\perp }^{2}+[k_{||}+A(t^{\prime
})]^{2}\right] }=0  \label{k1}
\end{equation}%
and%
\begin{equation}
\mathbf{k}_{\perp }^{2}\left( t-t^{\prime }+\frac{2i}{\left[ 2|E_{01}|+%
\mathbf{k}_{\perp }^{2}+[k_{||}+A(t^{\prime })]^{2}\right] }\right) =0,
\label{k2}
\end{equation}%
respectively. Apart from the trivial solution $\mathbf{k}_{\perp }=0,$ the
condition (\ref{k2}) can be satisfied by nonvanishing values of this
variable. One should note that, in the case without corrections, this does
not hold and only the trivial solution exists.

\begin{figure}[tbp]
\hspace*{-0.5cm}\includegraphics[width=9.2cm]{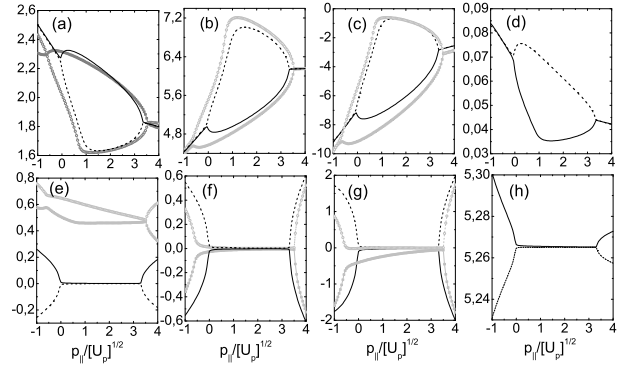}
\caption{Tunneling [Panels (a) and (e)] and rescattering times [Panels (b)
and (f)], together with the parallel [Panels (c) and (g)] and perpendicular
components of the intermediate momentum $\mathbf{k}$ [Panels (d) and (h)],
as functions of the parallel momentum $p_{||}$ along the diagonal $%
p_{1||}=p_{2||}$. The final transverse momenta $\mathbf{p}_{j\perp
}(j=1,2)$ are taken to be vanishing. The real and imaginary parts of
such quantities are displayed in the upper [(a), (b), (c) and (d)]
and lower [(e), (f), (g) and (h)] panels, respectively. The field
and atomic parameters are the same as in the previous figure. The
corrected and uncorrected yields are given by the black and gray
curves in the figure, respectively. The tunneling and rescattering
times are multiplied by $\protect\omega $ and the intermediate
momenta divided by $\protect\sqrt{\protect\omega }$, respectively,
so that a direct comparison with Figs. 5 and 6 can be performed.}
\end{figure}
Fig. 9 depicts the tunneling and rescattering times for this case,
together with the perpendicular and parallel components of
$\mathbf{k.}$ The real parts of such variables correspond, as in the
previous cases, to a longer and a shorter orbit. The momenta,
however, for which such orbits nearly coalesce, are radically
different from those in the previous cases discussed
in this paper. This is due to the fact that a nonvanishing $\mathbf{k}%
_{\perp }$ also affects the rescattering condition (\ref{saddle2}), which
now reads%
\begin{equation*}
\sum_{j=1}^{2}\left[ p_{j||}+A(t)\right] ^{2}=\left[ k_{||}+A(t)\right] ^{2}+%
\mathbf{k}_{\perp }^{2}-2|E_{02}|-\sum_{j=1}^{2}\mathbf{p}_{j\perp
}^{2}
\end{equation*}%
For constant final transverse momenta $\mathbf{p}_{j\perp }(j=1,2),$
this equation describes a circle centered at $-A(t)$ whose radius
has been altered in $\mathbf{k}_{\perp }^{2}$. Since, as shown in
Figs 9(a)-9(d), this radius decreased, $\mathbf{k}_{\perp }$ is
expected to be almost purely imaginary. This is indeed the case, as
can be seen comparing panels (d) and (h) in the figure. The
imaginary parts of such variables also behave following the same
pattern as previously, growing vary rapidly at the momenta for which
the real parts approach each other, and remaining nearly constant
in-between. Interestingly, $\mathrm{Im}[t^{\prime }]$ is vanishing
in this region. This feature is in clear contradiction with the fact
that tunneling is a process which is always forbidden, and therefore
requires a nonvanishing $\mathrm{Im}[t^{\prime }]$ (c.f. Fig.
5.(d)). For this reason, we will not use solutions with nonvanishing
$\mathbf{k}_{\perp }$ for computing electron momentum distributions.

\section{Influence of the ion}
\label{ion}

 In this section, we take a first step towards including
the residual ion in our formalism. For that purpose, we consider an
effective interaction $\widetilde{V}=V_{12}+V_{\mathrm{ion}}$ at the
time the first electron returns, where $V_{\mathrm{ion}}$ is the
ionic potential. Physically, this means that the first electron
interacts not only with the electron it releases, but, additionally,
with the residual ion. We take this potential to be of either
Coulomb or contact type, and assume that only the two active
electrons contribute to the ionic charge. Thus,
explicitly,$V_{\mathrm{ion}}$ reads
\begin{equation}
V^{(C)}_{\mathrm{ion}}=-2/|\mathbf{r}_1|\label{ioncoul}
\end{equation}
or
\begin{equation}
V^{(\delta)}_{\mathrm{ion}}\sim -2\delta
(\mathbf{r}_1).\label{ioncont}
\end{equation}
In this context, both the effective charge and a contact-type
interaction are justified by the fact that the remaining electrons
are screening the charge and the long-range tail of the binding
potential.

In Eq. (\ref{presc}), the form factors $V_{\mathbf{p_j},\mathbf{k}}$
are given by
\begin{equation}
\widetilde{V}^{(1s)}_{\mathbf{p}_{j},\mathbf{k}}\sim
V^{(1s)}_{\mathbf{p}_j,\mathbf{k}}
-\frac{2\eta(\mathbf{p}_1,\mathbf{k})}{[2|E_{02}|+(\mathbf{p}%
_{2}+\mathbf{A}(t))^{2}]^{2}}\newline
+(\mathbf{p}_{1}\leftrightarrow \mathbf{p}%
_{2}),\label{ionvpks}
\end{equation}

\begin{equation}
\widetilde{V}^{(2p)}_{\mathbf{p}_{j},\mathbf{k}}\sim V^{(2p)}_{\mathbf{p}_j,\mathbf{k}}-\frac{2\eta(\mathbf{p}_1,\mathbf{k})\sqrt{(\mathbf{p}_{2}+\mathbf{A}(t))^{2}%
}}{[2|E_{02}|+(\mathbf{p}_{2}+\mathbf{A}(t))^{2}]^{3}}\newline
+(\mathbf{p}_{1}\leftrightarrow \mathbf{p}_{2}),\label{ionvpkp}
\end{equation}
for $1s$ and $2p$ states, respectively. The prefactor $V_{12}$ is of
contact or Coulomb type. Furthermore, for a Coulomb or contact ionic
potential, $\eta$ is either constant or given by Eq.
(\ref{etacoul}), respectively. In order to simplify the
computations, and since only minor differences have been observed in
this case, we use the model in Sec. II, instead of the more rigorous
approach of Sec. IV in the subsequent figures.

Fig.~10 depicts how the ion affects the electron momentum
distributions, if its potential is assumed to be of Coulomb form
(Eq.(\ref{ioncoul})). The upper and lower panels have been computed
for $V_{12}$ of contact and Coulomb type, respectively. In the
figure, the distributions resemble those obtained for the
Coulomb-type interaction, if the second state is in a localized
state (Figs. 2.(d), 4.(d), and 8.(d)). This holds both for $1s$ and
$2p$ initial electron states. An inspection of Eqs. (\ref{ionvpks})
and (\ref{ionvpkp}) explains this shape. Indeed, in both equations,
the functional form of $\eta(\mathbf{p}_j,\mathbf{k})$, which is
characteristic of long-range interactions, favors unequal momenta,
leading to patterns similar to those observed in Figs. 2, 4, and 8.
Furthermore, in the second terms in
$\widetilde{V}_{\mathbf{p}_j,\mathbf{k}}$, the denominators are
small if $\mathbf{p}_j\simeq-\mathbf{A}(t)$. Thus, since, $A(t)\sim
2\sqrt{U_p}$, we expect the form factors (\ref{ionvpks}) and
(\ref{ionvpkp}) to be large near $p_{2||}=p_{1||}=\pm 2\sqrt{U_p}$.
Consequently, the yield in the diagonal gets enhanced. This is a
feature shared with the limit for localized wave functions, so that
the distributions are similar.
\begin{figure}[tbp]
\includegraphics[width=8.5cm]{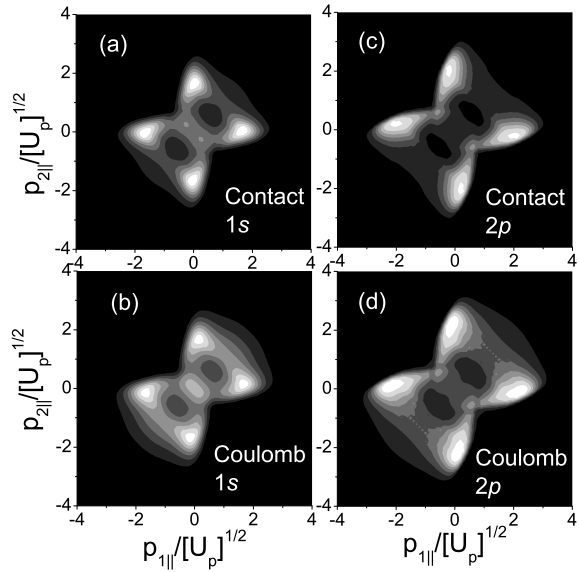}
\caption{Electron momentum distributions computed with the modified
form factors (\protect{\ref{ionvpks}}),(\protect{\ref{ionvpkp}}), as
functions of the electron momentum components parallel to the
laser-field polarization, for same field and atomic parameters as in
the previous figures. We consider that the ionic potential is of
Coulomb type (Eq. (\protect{\ref{ioncoul}})), and use the
saddle-point model without modifications (c.f. Sec. II). In the
upper and lower panels, the electron-electron interaction $V_{12}$
was assumed to be of contact and Coulomb type, respectively. In
panels (a), (b), both electrons are taken to be initially in $1s$
states, whereas in panels (c) and (d), they are initially in $2p$
states. The transverse momenta have been integrated over.}
\end{figure}

The subsequent figure (Fig. 11) is the counterpart of Fig. 10 for a
contact-type ionic potential (Eq. (\ref{ioncont})). In this case,
for all types of electron-electron interaction $V_{12}$ and initial
bound states, the distributions are strongly localized near
$p_{2||}=p_{1||}=\pm 2\sqrt{U_p}$, even though their shapes are
slightly different. This happens due to the fact that, in this case,
$\eta(\mathbf{p}_j,\mathbf{k})=const$. Therefore, the second terms
in the form factors (\ref{ionvpks}) and (\ref{ionvpkp}) are large
near $p_{2||}=p_{1||}=\pm 2\sqrt{U_p}$, but, in contrast to the
Coulomb-potential case, no unequal momenta are favored. This means
that the inclusion of a short-range ionic potential leads to a
radical improvement in the agreement between theory and experiment.
In this context, for both the contact- and Coulomb-type interactions
$V_{12}$, circular shapes reminiscent of those in the experiments
are only obtained if we consider $2p$ states. As previously
discussed, such states provide a more realistic description of the
outer-shell electrons in neon, as compared to $1s$ states.
\begin{figure}[tbp]
\includegraphics[width=8.5cm]{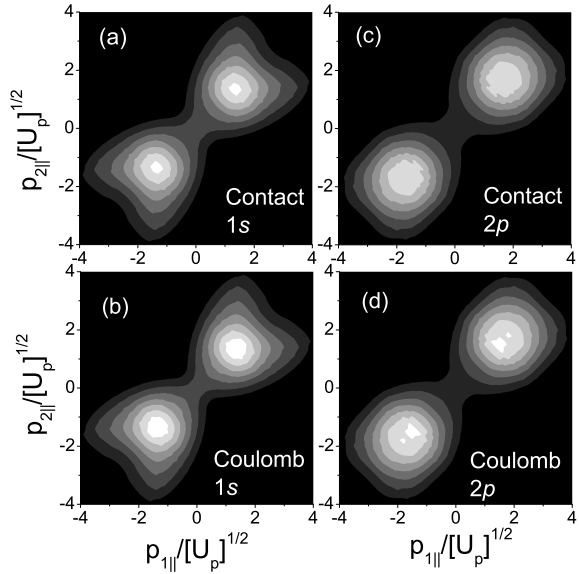}
\caption{Electron momentum distributions computed with the modified
form factors (\protect{\ref{ionvpks}}),(\protect{\ref{ionvpkp}}), as
functions of the electron momentum components parallel to the
laser-field polarization, for same field and atomic parameters as in
the previous figures. We consider that the ionic potential is of
contact type (Eq. (\protect{\ref{ioncont}})), and use the
saddle-point model without modifications (c.f. Sec. II). In the
upper and lower panels, the electron-electron interaction $V_{12}$
was assumed to be of contact and Coulomb type, respectively. In
panels (a), (b), both electrons are taken to be initially in $1s$
states, whereas in panels (c) and (d), they are initially in $2p$
states. The transverse momenta have been integrated over.}
\end{figure}

\section{Conclusions}
\label{concl}
 In this article, we have introduced several technical
 modifications in an S-Matrix theory of laser-induced nonsequential
 double ionization (NSDI), within the Strong-Field Approximation,
 in which this phenomenon is modeled as the inelastic collision
 of an electron with its parent ion. Such modifications include
 different initial bound states for the first and second electron,
 an adequate treatment of the bound-state singularity which exists
 in our framework, and an effective form factor
 which incorporates the residual ion. We performed a systematic analysis of
 their influence on the differential electron momentum
 distributions as functions of the parallel-momentum components
 $p_{j||}$, $(j=1,2)$ of both electrons.

 Specifically, we consider that the second electron is dislodged by a
 contact- and Coulomb-type interaction, and assume that both electrons are
initially in a $1s$, $2p$ or $3p$ hydrogenic state. As an additional
case, we assume that the first electron is initially bound in a $1s$
state, and that the initial wave function of the second electron is
localized at $\mathbf{r}_{2}=0.$ For the first electron, we take
into account only the hydrogenic states, since a neutral atom, in
contrast to a singly ionized atom, has a long-range binding
potential.

Concerning the initial bound-state wave function of the second
electron, our results show that the NSDI momentum distributions are
very sensitive to its spatial extension, but not to its shape.
Indeed, a spatially extended wave function causes a broadening in
the electron momentum distributions along the anti-diagonal
$p_{1||}=-p_{2||},$ even if the second electron is dislodged by a
contact-type interaction. Circular-shaped distributions, as
reported in \cite{pransdi1,pransdi2} and observed in experiments \cite%
{nsdireview,expe1} are only obtained for a contact-type interaction under
the additional condition that the bound-state wave function is localized at
the origin of the coordinate system, i.e., at $\mathbf{r}_{2}=0.$ In
addition to this broadening, if the second electron is released by a
Coulomb-type interaction, there is an enhancement in the contributions near
the axis $p_{1||}=0$ or $p_{2||}=0.$

All the distributions investigated in this article, however, change in a
less radical fashion if the second electron is taken to be in a
a $1s$, $2p$ or $3p$ hydrogenic state, as long as they exhibit a spatial
extension. In fact, although specific changes are observed, such as an
additional substructure for a Coulomb-type interaction, or more localized
distributions for a contact-type interaction, the overall shapes of such
distributions remains similar.

Furthermore, if the form factor $V_{\mathbf{k},0},$ which, within
our model, contains all the influence of the initial state of the
first electron, is incorporated in the time-dependent action, the
only noticeable effect is a suppression in the yield, for regions of
\ small parallel momenta. Indeed, the distributions retain their
shapes even if the saddle-point equations are modified in this way.
Such changes have been introduced in order to correct a singularity
which exists for such the prefactor $V_{\mathbf{k},0}$, within the
saddle-point framework, if the initial bound state is exponentially
decaying.

Finally, the inclusion of the ionic potential at the time of
rescattering, as the modified form factors
$\widetilde{V}_{\mathbf{p}_j,\mathbf{k}}$, sheds some light on why,
in the absence of the ion, a contact-type interaction localized at
the origin of the coordinate system yields the best agreement with
the experimental findings.

In fact, the ionic interaction leads to form factors which are very
large near $p_{1||}=p_{2||}=\pm \sqrt{U_p}$. This causes an
enhancement in the distributions in this region. If the ionic
potential is of Coulomb type, this effect is overshadowed by the
fact that $\eta(\mathbf{p}_j,\mathbf{k})$, given by Eq.
(\ref{etacoul}), favors unequal momenta. By contrast, if the ionic
potential is given by Eq. (\ref{ioncont}), which is a good
approximation for a short-range interaction,
$\eta(\mathbf{p}_j,\mathbf{k})=const.$ and the enhancement at the
diagonal prevails. On the other hand, in Sec. III and IV, if
$V_{\mathbf{p}_j,\mathbf{k}}=const.$ (i.e., for $V_{12}$ of
contact-type and a localized state for the second electron), the
very same effect is caused by integrating over the phase space.
Interestingly, if, in the presence of the ion, we consider $2p$
states, which are more realistic assumptions for our model, the
agreement with the experimental findings improves even more.

In conclusion, the present results indicate that the ionic potential
 is an important ingredient for a realistic modeling of NSDI. Indeed,
 of all the technical modifications considered in this
paper, which aimed at making the model more realistic, this was the
only which played a major role in improving the agreement between
theory and experiment. The other modifications either worsened this
agreement, or had almost no influence on the momentum distributions.
This supports the hypotheses raised in previous studies
\cite{pransdi1,pransdi2}, that the residual ion might be screening
both the long-range of the Coulomb interaction, or the final-state
Coulomb repulsion, so that, effectively, the electron-electron
interaction is of contact-type, and the bound-state wave functions
are localized.

We would like to stress out, however, that the treatment performed
in Sec. V is only a first approximation for a rigorous study of the
ionic potential. There exist, in principle, more rigorous methods
for incorporating the residual ion. The first approach would be to
consider the ion as a further interaction in our model, and modify
the transition amplitude accordingly. This is, however, a highly
non-trivial task, since it would lead to one more rescattering and a
further integral in the transition amplitude. Another possibility
would be to incorporate the ionic potential in the propagation of
both electrons in the continuum. This would allow a clear assessment
of the Coulomb focusing, which, again, owes its existence to the
presence of the ion. Indeed, this effect may as well be compensating
the broadening caused by initial spatially extended wave functions.
Definite statements on this issue, however, require a theoretical
approach beyond the Strong-Field Approximation.

\begin{acknowledgments}
This work was financed in part by the Deutsche
Forschungsgemeinschaft (SFB 407 and European Graduate College
``Interference and Quantum Applications"). We are grateful to A.
Sanpera for her collaboration in the initial stages of this project,
and to one of the referees for pointing out the possible role of the
ionic potential. C.F.M.F. would like to thank the Optics Section at
the Imperial College and City University for their kind hospitality,
and W. Becker for useful discussions.
\end{acknowledgments}

\end{document}